\documentclass[onecolumn,12pt,nobibnotes,showpacs,showkeys]{revtex4}

\usepackage{graphicx}
\usepackage{amsmath}
\usepackage[dvips]{color}
\usepackage{newlfont}
\usepackage[latin1]{inputenc}
\usepackage{float}

\begin{document}

\title{Effectively irreversible fringe contrast attenuation induced by one degree of freedom}

\begin{abstract}

The attenuation of fringe contrast in a Ramsey interferometer induced by the atom-field interaction is analyzed.
We show that short-time power series expansion is not the proper tool to find the relevant time scale for such process.
Analytical expressions to quantify the relevant time scale for the fringe contrast decay and to characterize the long-term effectiveness of this process are proposed.
For Pegg-Barnett phase state initial conditions,
these expressions suggest that: the increase of the energy of the
field leads to slower vanishing of fringe contrast; the increase
of the field photon number variance leads to more effective
attenuation in the final regime. 
Numerical simulations with coherent and thermal initial states are in qualitative agreement with such results.
\end{abstract}

\author{A. R. Bosco de Magalhães}
\affiliation{Departamento de F{í}sica e Matemática, Centro Federal
de Educa{çã}o Tecnológica de Minas Gerais, Belo Horizonte, MG,
30510-000, Brazil }
\author{Adélcio C. Oliveira}
\affiliation{Departamento de F{í}sica e Matemática, Universidade Federal de São João Del Rei, C.P. 131,
Ouro Branco, MG, 36420-000, Brazil }
\keywords{two-way interferometry, classical limit, cavity quantum electrodynamics}
\pacs{03.65.Yz, 03.65.Aa, 42.50.Pq}
\date{\today} 
\maketitle

\section{Introduction}

\label{I}

The deleterious action of the environment over quantum coherences has been a
fundamental ingredient in the study of the foundations of quantum mechanics,
for it sheds light on the quantum to classical transition problem \cite
{Giulini}. This process is called decoherence, and also plays a central role
in quantum information, as it is the main obstacle for quantum computation 
\cite{Bouwmeester}. 
Progressive loss of coherence was observed in an important cavity quantum
electrodynamics (QED) experiment \cite{Brune1}.
At the theoretical level, decoherence is frequently
studied by considering that the system of interest is coupled to another
system, the environment \cite{Senitzky}: if we take the trace over the
environmental degrees of freedom, we access the statistics we are interested in. 
Other decoherence approaches are presented in Ref. \cite{Kofler}, based on
coarse-grained measurements, and in Ref. \cite{Bonifacio}, where a
fluctuation of some classical parameter is responsible for coherence loss.

Although the environment is usually modeled as a many degrees of freedom system 
\cite{Ford1,Ford2,Caldeira}, decoherence has also been investigated by
coupling the system of interest to few degrees of freedom
whose classical analogs exhibit chaotic (or chaotic-like)
behaviors \cite{Furuya,Blume-Kohout2,Rossini,Casati,Bandyopadhyay,Lemos}. 
In Ref. \cite{Oliveira1}, we used a quartic oscillator coupled through cross-Kerr
interaction to a variable number of bosons to show that 
effectively irreversible loss of quantum coherences may be induced 
even by one degree of freedom environment without chaos.
In the present contribution, we continue the research on these lines by focusing on the
effectively irreversible attenuation of fringe contrast in a two-way interferometer
induced by one degree of freedom.
The implementation of such an interferometer in the cavity QED context is described in Ref. \cite{Bertet}.
That implementation is particularly interesting for the present discussion,
since the single field mode that fulfills the function of a beam splitter 
also plays a role analogous to the one often played by infinity degrees of freedom reservoirs:
due to the atom-field entanglement, it makes non-diagonal elements of the reduced atomic density operator to decrease, leading to diminution of the fringe contrast. 
In order to study the interferometer in Ref. \cite{Bertet}, we naturally employed the Jaynes-Cummings model \cite{Cummings} in the rotating wave approximation, as addressed in Section \ref{M}. 
We wish to emphasize that although this model has been widely studied in the past decades,
we believe our results are new, since the Jaynes-Cummings model is used here as a means to study the induction of classicality by few degrees of freedom, which has attracted attention especially in recent times. 

The characterization of relevant time scales for the fringe contrast decay (FCD) requires, in the present case, a different approach from the one in Ref. \cite{Oliveira1}, where decoherence time scales were defined by means of short-time power series expansions, as usual \cite{Kim}. Thus, in Section \ref{AR} we propose analytical expressions based on dephasing of complex terms to quantify FCD time scales. There, we also indicate a method to quantify the effectiveness of the attenuation process for times much longer than the calculated time scales.
These results, obtained for Pegg-Barnett phase state \cite{Pegg} bosonic initial conditions,
suggest that: 
a) increasing the photon number variance of the field leads to more effective
attenuation of the fringe contrast in the final regime; 
b) increasing the energy of the field leads to slower vanishing of such contrast.
In Section \ref{SI}, we focus on the
Ramsey interferometer described in Ref. \cite{Bertet}.
We show that for large mean photon number the interference fringes would disappear if
long atom-field interaction times were taken. This agrees qualitatively with the results in 
Section \ref{AR}.
In Section \ref{SDR}, we analyze the thermal initial bosonic state case
by taking into account the parameters encountered in a recent experiment \cite{Gleyzes}.
Our conclusions are drawn in Section \ref{C}.

\section{The model}

\label{M}

Let us consider a two level system coupled to one resonant oscillator by the
Hamiltonian 
\begin{equation}
\mathbf{H}=\frac{1}{2}\hbar \omega \mathbf{\sigma }_{z}+\hbar \omega \mathbf{%
a}^{\dagger }\mathbf{a}+\hbar g\left( \mathbf{\sigma }_{+}\mathbf{a}+\mathbf{%
\sigma }_{-}\mathbf{a}^{\dagger }\right) ,  \label{spinbosonhamiltonian}
\end{equation}
where $\mathbf{a}^{\dagger }$ and $\mathbf{a}$ are creation and annihilation
bosonic operators, and $\mathbf{\sigma }_{z}=\left| e\right\rangle
\left\langle e\right| -\left| g\right\rangle \left\langle g\right| $, $%
\mathbf{\sigma }_{+}=\left| e\right\rangle \left\langle g\right| $ and $%
\mathbf{\sigma }_{-}=\left| g\right\rangle \left\langle e\right| $ are
spin-1/2 operators. This Hamiltonian may concern a Rydberg atom coupled to
one microwave mode in a lossless superconducting cavity \cite{Raimond2}. 
Although analytically solvable, the Jaynes-Cummings model presents a rich dynamics. 
This is unveiled in seminal papers about the collapses and revivals 
of the atomic population inversion \cite{Eberly,Narozhny,Yoo}.
Noteworthy results are also displayed in Ref. \cite{Phoenix1}, which focus on the behavior of the entropy in the collapse region of the inversion. 
Other remarkable contributions in the 
long history of this model are found in Refs. \cite{Gea-Banacloche1,Gea-Banacloche2}.
There, it was shown that the atomic state approaches a pure state (independent of
the initial state of the atom) in the middle of the collapse region: this
was interpreted as a collapse of the wave function, since the information
about the initial atomic state was transfered to the (macroscopic) field.
Such a phenomenon was also studied in Ref. \cite{Phoenix2}, where the
atom-field state was given in terms of the eigenvalues and eigenstates of
the reduced atomic and field density operators. 

\section{Analytical results}

\label{AR}

We begin by analyzing the dynamics for the initial state 
\begin{equation}
\left| \psi \left( 0\right) \right\rangle 
=\left( c_{e}\left| e\right\rangle +c_{g}\left|
g\right\rangle \right) \otimes \underset{n=0}{\overset{r}{\sum }}\frac{1}{%
\sqrt{r+1}}\left| n\right\rangle .
\label{rho(0)r}
\end{equation}
The bosonic state is a particular approximate
Pegg-Barnett phase state. 
This is not usually built in cavity fields, but it permit us to get
analytical results. 

\begin{figure}[th]
\vspace{0cm} \hspace{0cm} \includegraphics[scale=0.7]{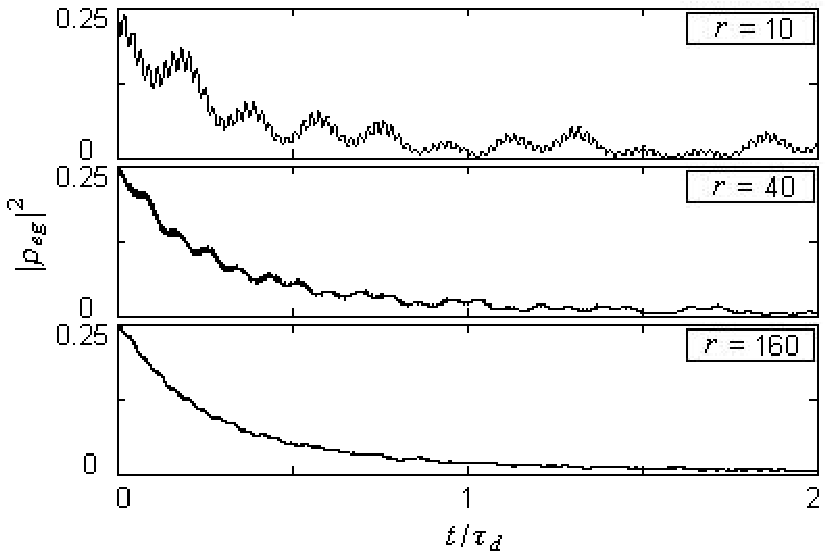} 
\vspace{0cm}
\caption{Time evolution of $\left| \protect\rho _{eg}\right| ^{2}$ for
initial state (\ref{rho(0)r}) with $c_{e}=c_{g}=1/\sqrt{2}$.}
\label{figp1}
\end{figure}

The fringe contrast in the Ramsey interferometer of Ref. \cite{Bertet} is proportional to   
the absolute value of the non-diagonal elements of the reduced atomic density operator
\begin{equation}
\rho _{a} =\rho _{ee} \left| e\right\rangle
\left\langle e\right| +\rho _{gg} \left| g\right\rangle
\left\langle g\right| +\rho _{eg} \left| e\right\rangle
\left\langle g\right| +\rho _{ge} \left| g\right\rangle
\left\langle e\right|
\end{equation}
right after the resonant atom-field interaction.
Let us first consider $c_{e}=c_{g}=1/\sqrt{2}$, what leads to
\begin{equation}
\rho _{eg}\left( t\right) =\frac{1}{2\left( r+1\right) }
\left\{ i e^{ig \sqrt{r} t} \sin \left( g \sqrt{r+1} t \right)
+\underset{n=0}{\overset{r}{\sum }}e^{-ig\left( \sqrt{n+1}-%
\sqrt{n}\right) t}\right\} .  
\label {rhoegr}
\end{equation}
This matrix element is calculated as a sum of
complex terms with different phases. When $t=0$, the phases are correlated
(all the terms are real) and $\left| \rho _{eg}\right| $ assumes its maximum
value. A FCD time scale may be defined as the time required for 
the vanishing of most of these correlations. Such time will be assumed here as
the one spent by the slowest
term in Eq. (\ref{rhoegr}) to make a complete oscillation: 
\begin{equation}
\tau _{d}=\frac{2\pi }{g\left( \sqrt{r+1}-\sqrt{r}\right) }, 
\label{td}
\end{equation}
which grows with $\sqrt{r}$ for large $r$.
The evolutions of $\left| \rho_{eg}\right| ^{2}$ plotted in Fig. \ref{figp1} 
corroborate the validity of Eq. (\ref{td}). 
This behavior was observed for $r\leq160$, larger r could not be
investigated due to numerical limitations.

\begin{figure}[th]
\vspace{0cm} \hspace{0cm} \includegraphics[scale=0.7]{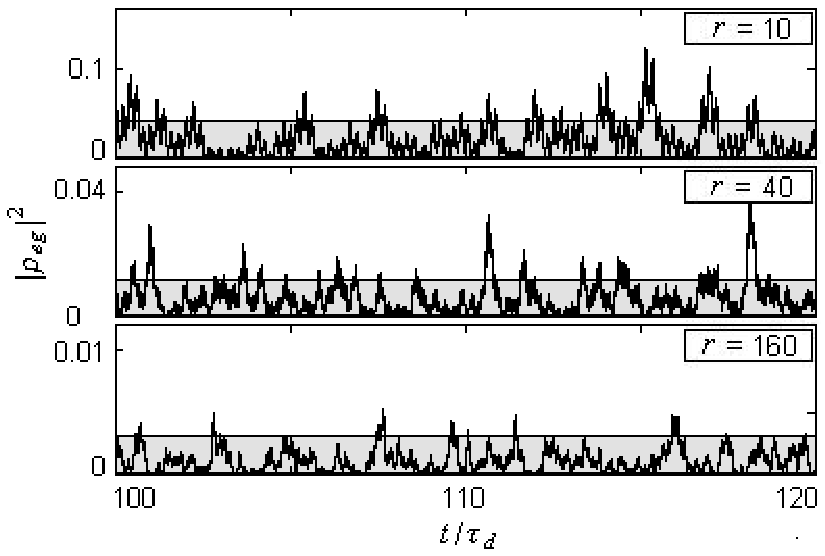} 
\vspace{0cm}
\caption{Time evolution of $\left| \protect\rho _{eg}\right| ^{2}$ for
initial state (\ref{rho(0)r}) with $c_{e}=c_{g}=1/\sqrt{2}$. The shading region corresponds to values of $%
\left| \protect\rho _{eg}\right| ^{2}$ between $M-\protect\sigma $ and $M+\protect\sigma $.}
\label{figp2}
\end{figure}

When $t=\tau _{d}$, most of the phase correlations of the terms in Eq. (\ref{rhoegr}) is lost: the low value
of $\left| \rho _{eg}\left( \tau _{d}\right) \right| $ is a consequence of
the mutual cancellation of the terms in the sum. Although revivals of
$\left| \rho _{eg}\right| $ may occur, they will not be complete, since the
frequencies of the periodic functions summed in Eq. (\ref{rhoegr}) are not
commensurable. If $r$ grows, each term in Eq. (\ref{rhoegr}) gets smaller
and the mutual cancellation becomes more effective.
Along these lines we propose a
way to quantify such process: assuming that for long times the complex
phases (and also the argument in the sine function) behave as random
variables, with no correlation among them, uniformly distributed between $0$
and $2\pi $, the mean value and the standard deviation of $\left| \rho
_{eg}\right| ^{2}$ will be respectively given by 
\begin{align}
M& =\left( \frac{1}{2\left( r+1\right) }\right) ^{2}\left( r+\frac{1}{2}%
\right) ,  \label{media} \\
\sigma & =\left( \frac{1}{2\left( r+1\right) }\right) ^{2}\sqrt{r^{2}+\frac{1%
}{8}},  \label{variancia}
\end{align}
decreasing with $1/r$ for large $r$. 
The long time behavior of $\left| \rho _{eg}\right| ^{2}$
is displayed in Fig. \ref{figp2},
where the effectiveness of Eqs. (\ref{media}) and (\ref{variancia})
is exemplified in different orders of magnitude (notice the different vertical axes scales).

\begin{figure}[th]
\vspace{0cm} \hspace{0cm} \includegraphics[scale=0.7]{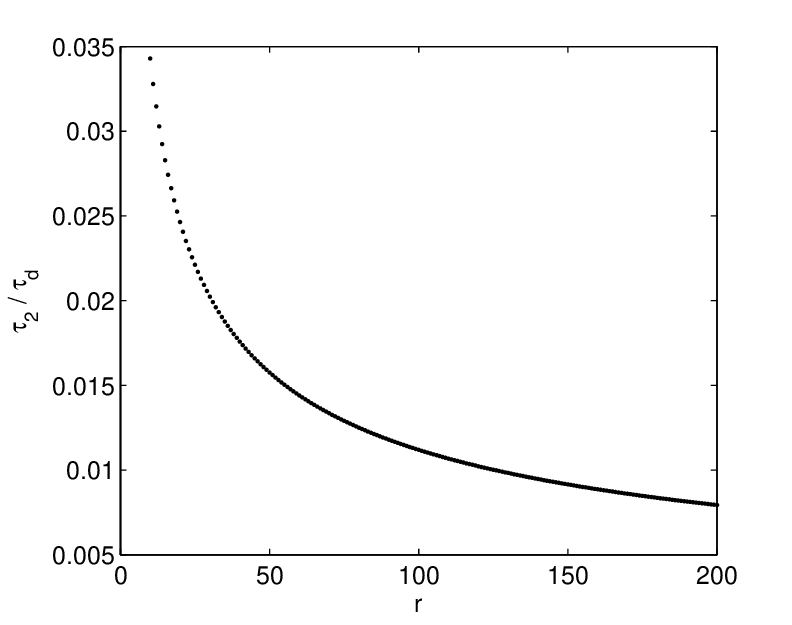} 
\vspace{0cm}
\caption{Ratio $\tau_{2} / \tau_{d}$ for initial state (\ref{rho(0)r}) with $c_{e}=c_{g}=1/\sqrt{2}$.}
\label{pst2td}
\end{figure}

\begin{figure}[th]
\vspace{0cm} \hspace{0cm} \includegraphics[scale=0.7]{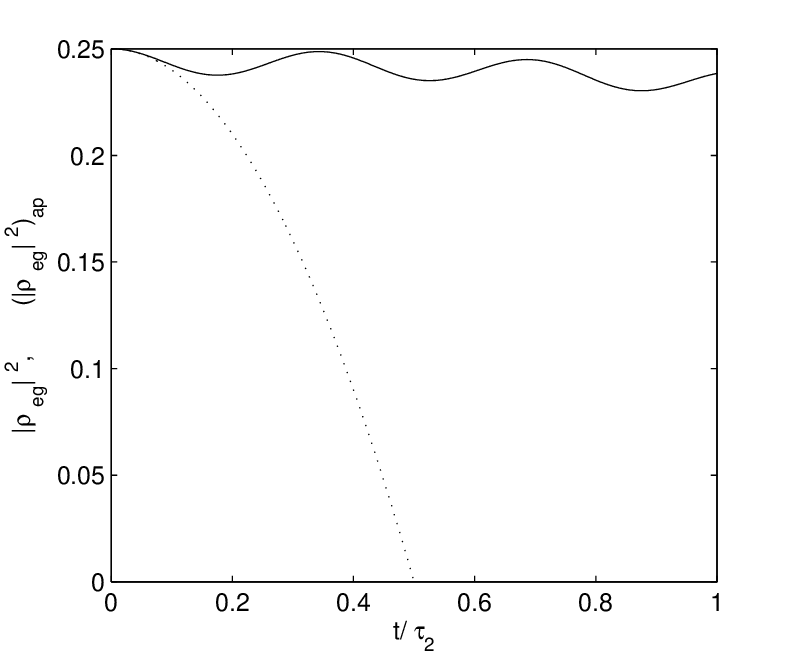} 
\vspace{0cm}
\caption{Time evolution of $\left| \protect\rho _{eg}\right| ^{2}$ (solid line) 
and approximation (\ref{rhoegaprox}) (dotted line) for
initial state (\ref{rho(0)r}) with $r=40$ and $c_{e}=c_{g}=1/\sqrt{2}$.}
\label{figtau2}
\end{figure}

Let us look at the relevance of calculating a FCD time scale through short-time power series expansions for the present case. The Taylor series of $\left| \rho _{eg}\left( t\right) \right| ^{2}$ may be written as
\begin{equation} 
\left| \rho _{eg}\left( t\right) \right| ^{2}=\left| \rho _{eg}\left( 0\right) \right| ^{2}+\left( \left. \frac{d}{dt}\left( \left| \rho _{eg}\left( t\right) \right| ^{2}\right) \right| _{t=0}\right) t+\left( \frac{1}{2}\left. \frac{d^{2}}{dt^{2}}\left( \left| \rho _{eg}\left( t\right) \right| ^{2}\right) \right| _{t=0}\right) t^{2}+O\left( t^{3}\right) ,
\end{equation}
where $O\left( t^{3}\right) $ denotes the terms of third order in $t$. The first order term in this expansion is zero. Accordingly, 
\begin{equation}
\left| \rho _{eg}\left( t\right) \right| ^{2}\approx \left| \rho _{eg}\left( 0\right) \right| ^{2}+\left( \frac{1}{2}\left. \frac{d^{2}}{dt^{2}}\left( \left| \rho _{eg}\left( t\right) \right| ^{2}\right) \right| _{t=0}\right)
t^{2} = \left( \left| \rho _{eg}\left( t\right) \right| ^{2} \right) _{ap}
\label{rhoegaprox}
\end{equation}
for short times. The time scale for relevant changes in this approximation may be calculated as the time for the absolute value of the second order term to reach one, leading to the definition of the second order FCD time scale:
\begin{equation}
\tau _{2}=\frac{1}{\sqrt{\frac{1}{2}\left| \left. \frac{d^{2}}{dt^{2}}\left( \left| \rho _{eg}\left( t\right) 
\right| ^{2}\right) \right| _{t=0}\right| }}=\frac{2\sqrt{r+1}}{g\sqrt{2\sqrt{r\left( r+1\right) }+\underset{n=0}{\overset{r}{\sum }}\left( \sqrt{n+1}-\sqrt{n}\right) ^{2}}}.
\label{tau2}
\end{equation}
In Fig. \ref{pst2td}, we show the ratio $\tau _{2}/\tau _{d}$ for $10\leq r\leq 200$. It is clear, from the dynamics shown in Fig. \ref{figp1}, that the time scale for the decrease of $\left| \rho _{eg}\right| ^{2}$ is really given by $\tau _{d}$, at least for $r$ around the values investigated. In view of the very low ratios displayed in Fig. \ref{pst2td}, we conclude that $\tau _{2}$ does not give the relevant time scale for the vanishing of the fringe contrast. This may be understood with the help of Fig \ref{figtau2}, where we compare $\left| \rho _{eg}\left( t\right) \right| ^{2}$ with its second order approximation for $r=40$. We see that although the approximation (\ref{rhoegaprox}) is good for very short times, the higher order terms become relevant long before the second order term reaches one. Namely, the second order approximation is not good in the whole interval $0\leq t\leq \tau _{2}$, what turns the definition (\ref{tau2}) uncorrelated to the decay of the fringe contrast.

\begin{figure}[th]
\vspace{0cm} \hspace{0cm} \includegraphics[scale=0.7]{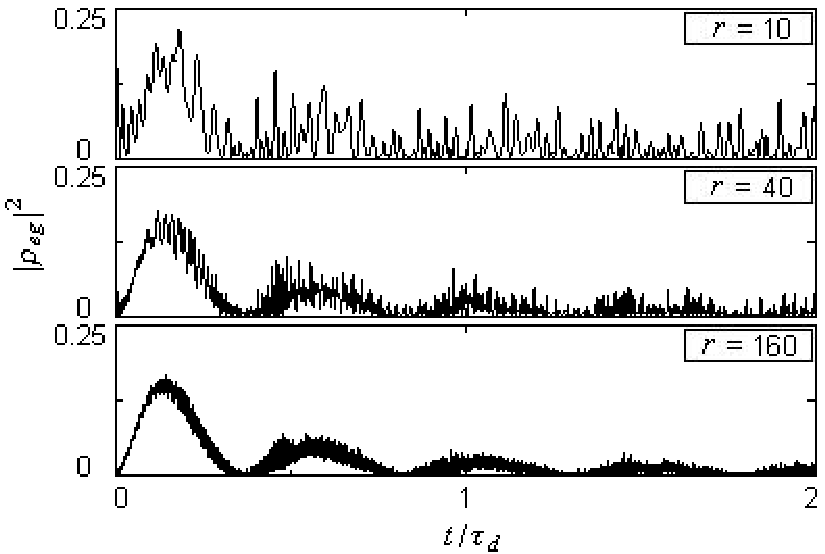}
\vspace{0cm}
\caption{Time evolution of $\left| \protect\rho _{eg}\right| ^{2}$ for
initial state (\ref{rho(0)r}) with $c_{e}=1$ and $c_{g}=0$.}
\label{ps3}
\end{figure}

\begin{figure}[th]
\vspace{0cm} \hspace{0cm} \includegraphics[scale=0.7]{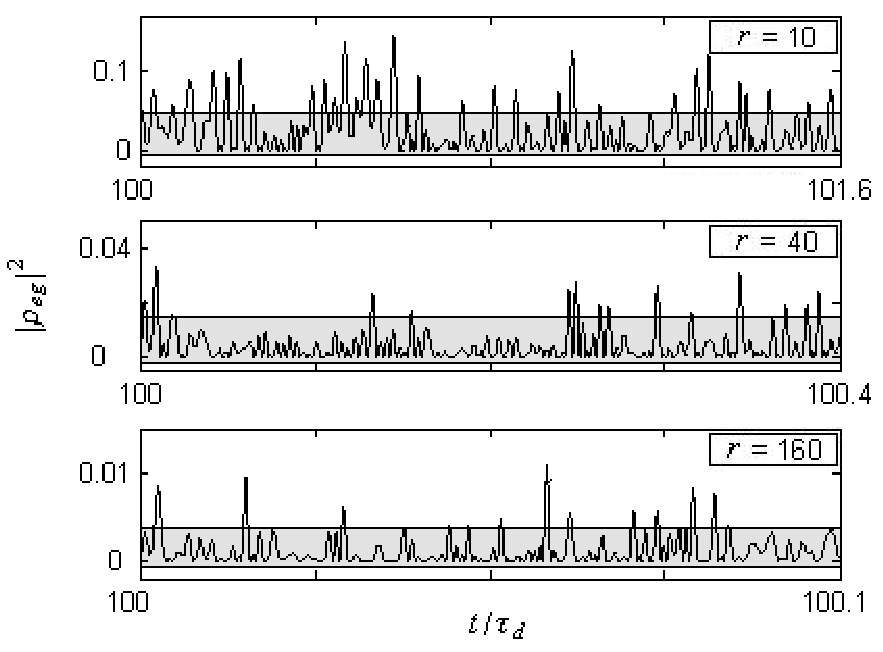} 
\vspace{0cm}
\caption{Time evolution of $\left| \protect\rho _{eg}\right| ^{2}$ for
initial state (\ref{rho(0)r}) with $c_{e}=1$ and $c_{g}=0$. The shading region corresponds to values of $%
\left| \protect\rho _{eg}\right| ^{2}$ between $M-\protect\sigma $ and $M+%
\protect\sigma $. }
\label{ps4}
\end{figure}

Related results are obtained for $c_{e}=1$ and $c_{g}=0$. 
This is the phase state case corresponding to the coherent state case analyzed in Ref. \cite{Bertet}.
Now, $\left| \rho _{eg} \right|^{2} $ starts at zero and performs damped slow oscillations 
with characteristic time scales calculated as above (see Fig. \ref{ps3}).
The evolution of $\rho _{eg} $ may be displayed in the form
\begin{equation}
\rho _{eg}\left( t\right) =\frac{i}{2\left( r+1\right) }\underset{n=1}{%
\overset{r}{\sum }}\left\{ \sin \left[ g\left( \sqrt{n}-\sqrt{n+1}\right)
t \right]+\sin \left[ g\left( \sqrt{n}+\sqrt{n+1}\right) t \right] \right\},
\label{rhoegre}
\end{equation}
and the time for the slowest term in Eq. (\ref{rhoegre}) to perform a complete oscillation defines the
characteristic time scale, which is also given by Eq. (\ref{td})
(the validity of this time scale is exemplified in Fig. \ref{ps3}).
Under random variables assumption for the arguments of the sine functions in Eq. (\ref{rhoegre}),
the mean value and the standard deviation of $\left| \rho _{eg}\right| ^{2}$ are given by 
\begin{align}
M& =\left( \frac{1}{2\left( r+1\right) }\right) ^{2}r,  \label{mediae} \\
\sigma & =\left( \frac{1}{2\left( r+1\right) }\right) ^{2}\sqrt{2r^{2}-\frac{%
15}{4}r+3}.  \label{varianciae}
\end{align}
The effectiveness of Eqs. (\ref{mediae}) and (\ref{varianciae})
is exemplified for different orders of magnitude in Fig. \ref{ps4},
where the plots were constructed with distinct vertical axes.
These plots were also constructed with distinct horizontal axes,
in order to display the relevant scale of the dynamics in the long time regime,
when the slow oscillations that decay with $\tau_{d}$ are no longer appreciable.
For each plot of Fig. \ref{ps4}, the faster terms of Eq. (\ref{rhoegre}) 
perform a number of complete oscillations between
$60$ and $70$ in the period shown.

In our numerical investigations we did not find any relevant revival where the behavior departs considerably from that shown in Figs. \ref{figp2} and \ref{ps4}. This suggests that we approach the irreversibility when $r$ increases.

\section{The Ramsey interferometer}

\label{SI}

In Ref. \cite{Bertet}, it is reported a complementarity experiment exploring interesting features of a
Ramsey interferometer. A Rydberg atom with relevant levels $e$
and $g$ is sent through a microwave cavity. The atom is initially in the
excited level $e$ and the field is prepared in the coherent state $\left| \alpha \right\rangle$.
Atom and field interact resonantly during a period $%
t_{\alpha }$ defined by $\rho _{ee}\left( t_{\alpha }\right) =1/2$. Then, by
applying an electric field across the cavity mirrors, the relative phase $%
\phi $ of the probability amplitudes related to levels $e$ and $g$ is
shifted by a variable amount. Finally, the atom crosses a Ramsey zone, and
the transition probability between levels $e$ and $g$ ends with the value 
\begin{equation}
P_{g}\left( \phi \right) =\frac{1}{2}\left\{ 1+Re\left( 2\rho
_{ge}\left( t_{\alpha }\right) \exp \left( i\phi \right) \right) \right\} .
\end{equation}
The contrast of the fringes depends on the atom-field entanglement \cite{Terra},
which is related to which-path
information and is responsible for diminishing $\left| \rho _{ge}\left(
t_{\alpha }\right) \right| $.

\begin{figure}[th]
\vspace{0cm} \includegraphics[scale=0.7]{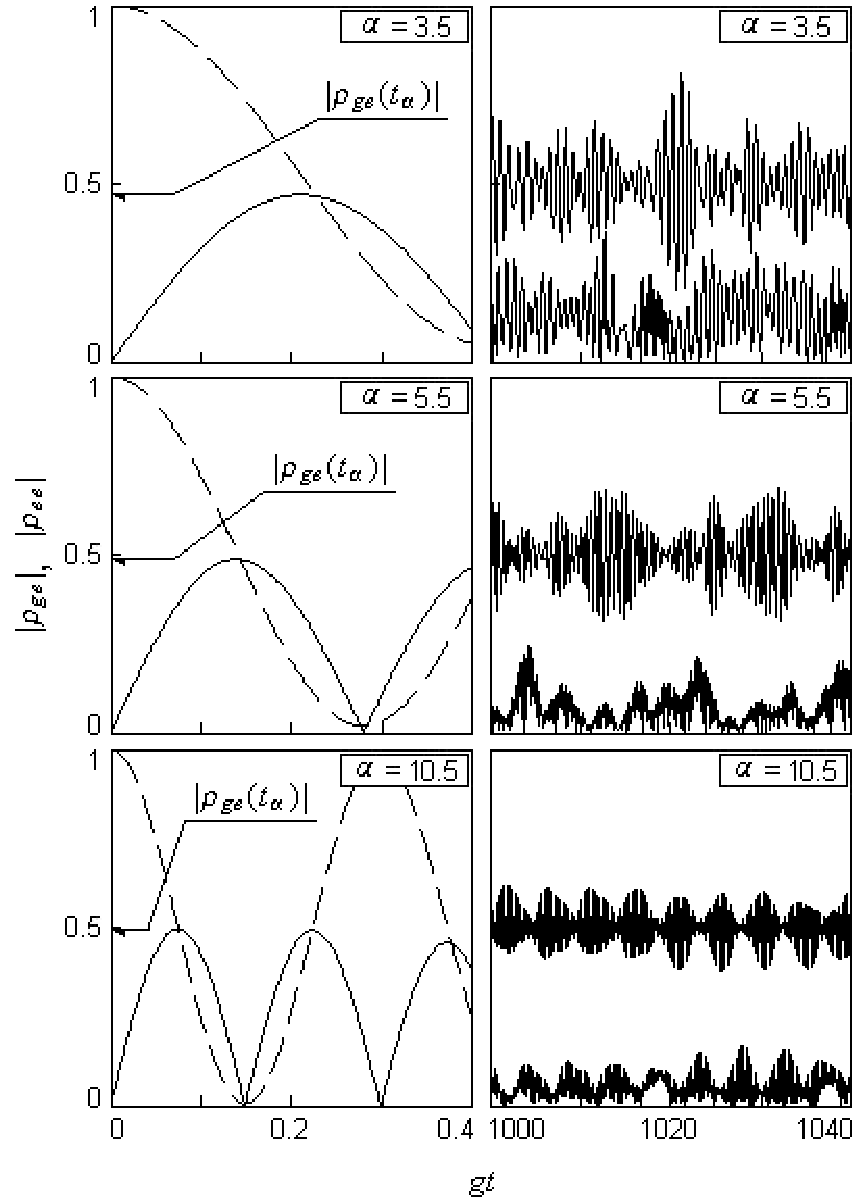} \vspace{0.0cm}
\caption{Time evolution of $\left| \protect\rho _{ge}\right| $ and $\left| 
\protect\rho _{ee}\right| $ for initial state $\protect\rho \left( 0\right)
=\left| e\right\rangle \left\langle e\right| \otimes \left| \protect\alpha %
\right\rangle \left\langle \protect\alpha \right| $. In the left plots,
solid lines correspond to $\left| \protect\rho _{ge}\right| $ and dashed
lines to $\left| \protect\rho _{ee}\right| $. In each right plot, lower
curves correspond to $\left| \protect\rho _{ge}\right| $ and upper curves to 
$\left| \protect\rho _{ee}\right| $. }
\label{figc3}
\end{figure}

The left plots of Fig. \ref{figc3} show the values for $\left| \rho
_{ge}\left( t_{\alpha }\right) \right| $ concerning the first time
$\rho _{ee}\left( t\right) $ reaches $1/2$. As in the experimental situation
in Ref. \cite{Bertet}, such values increase with $\left| \alpha \right| $:
the more the coherent field approaches a classical regime, the less
which-path information will be available on it. However, if it had been
chosen a much longer $t_{\alpha }$ (also satisfying $\rho _{ee}\left(
t_{\alpha }\right) =1/2$), an opposite behavior would be observed: $\left|
\rho _{ge}\left( t_{\alpha }\right) \right| $ would tend to be small for
high values of $\left| \alpha \right| $, as it is exemplified in the right
plots of Fig. \ref{figc3}. When $\left| \alpha \right| $ increases, the
field takes a longer time to store which-path information, but the flow back
of this information becomes less significant.

\begin{figure}[th]
\vspace{0cm} 
\includegraphics[scale=0.7]{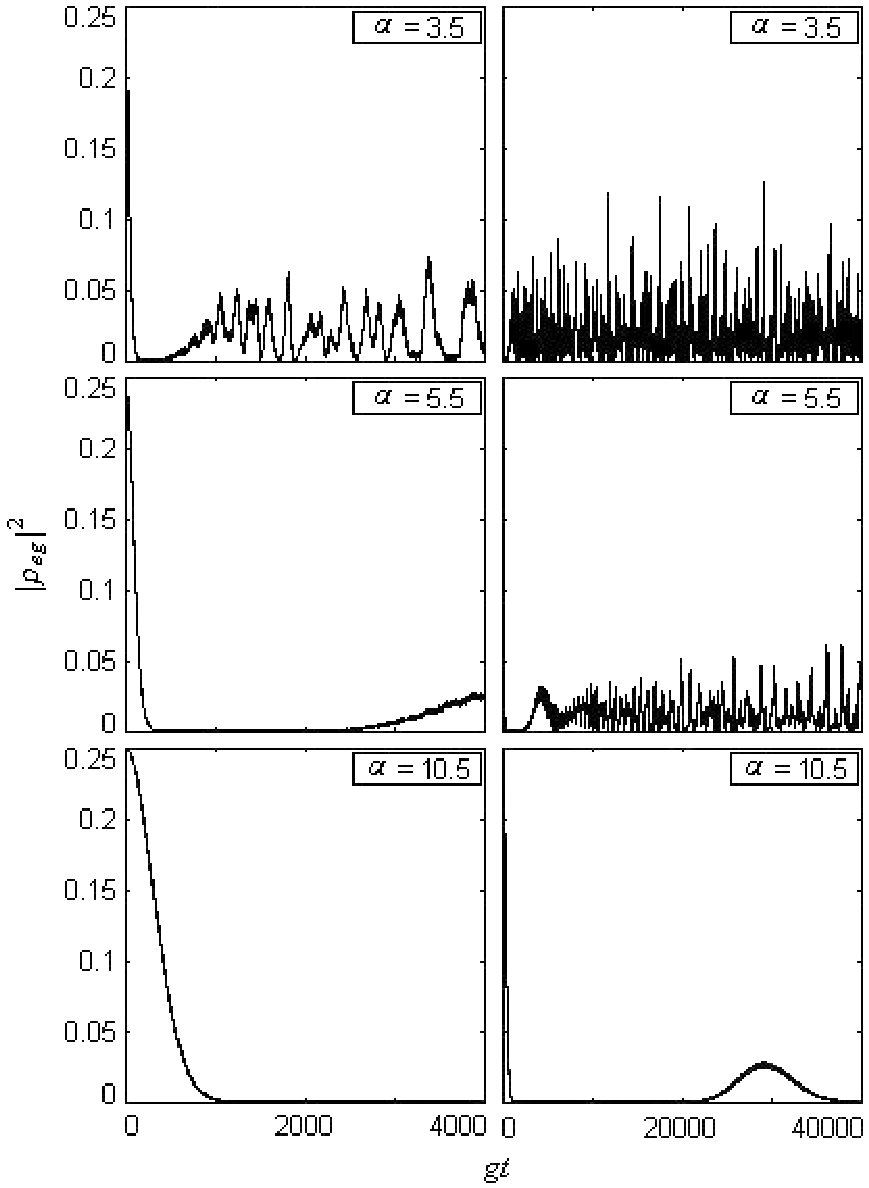} \vspace{0cm}
\caption{Time evolution of $\left| \protect\rho _{eg}\right| ^{2}$ for
initial state 
$\protect\rho \left( 0\right)=
\left( 1/2 \right)
\left( \left| e\right\rangle +\left|
g\right\rangle \right) \left( \left\langle e\right|
+\left\langle g\right| \right) \otimes \left| \protect\alpha %
\right\rangle \left\langle \protect\alpha \right| $.}
\label{figc}
\end{figure}

Similarities with the results in Section \ref{AR} become clear
with the help of Fig. \ref{figc}, which must be compared to Fig. \ref{figp1}.
Enhancing $\left| \alpha \right| $ or $r$ corresponds to
enhancing the mean photon number and the photon number variance. This leads,
in both cases, to more effective, but retarded, fringe contrast decay.
For the phase state case, such a retardation seems to be mainly
related to the energy increasing, since the slowest terms in expressions (\ref
{rhoegr}) and (\ref{rhoegre}), which determine the FCD time scale, correspond to the highest
occupied bosonic energy levels. 
The maintenance of the fringe contrast is related to the separability of the atom-field state.
Thus, this slowdown of the contrast decreasing resembles results found in Ref. \cite{Gea-Banacloche2}, 
where it is shown, for specific atomic initial states and
the field starting in a coherent state in the limit of large average photon
number $\bar{n}\longrightarrow\infty$, that the atom and field remain separately in a pure state,
for any finite time and even for infinite times provided that the time $t$ go to infinity slowly enough so that
$t/\bar{n}\longrightarrow 0$.
On the other hand, also for the phase state case, the long-term
effectiveness of the fringe contrast decay is mainly related to the photon number
variance: if this variance increases, each term in expressions (\ref{rhoegr}) and (\ref{rhoegre})
decreases, leading to more effective mutual cancellation due to the pseudo-randomization
of the relative phases.
This is reflected in the factors $1/\left( r+1\right) $ in Eqs. (\ref{mediae}) and (\ref{varianciae}). 
Although all energy levels are present in the coherent state,
relevant coefficients are found in a finite range,
and a similar mechanism for FCD process may exist.

\section{Taking recent experimental parameters into account}

\label{SDR}

\begin{figure}[th]
\vspace{0cm} \includegraphics[scale=0.7]{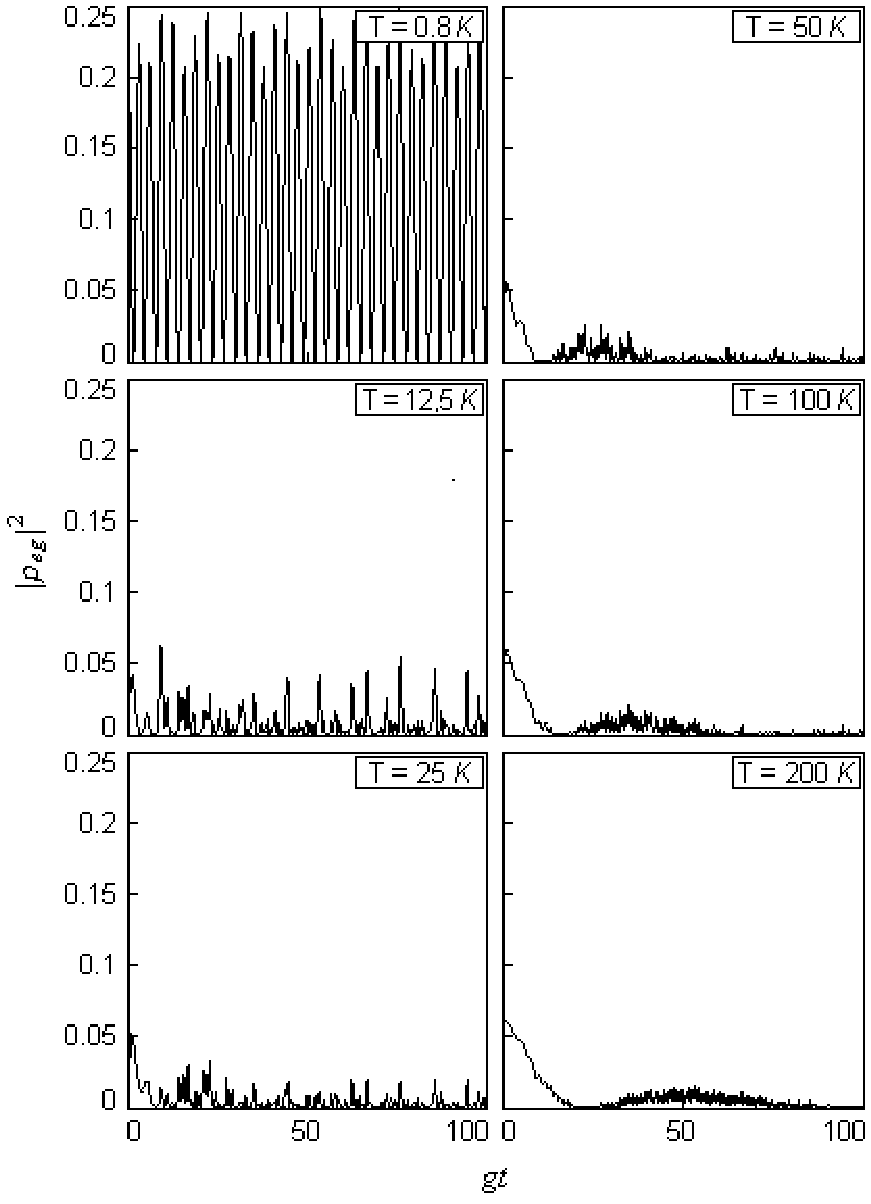} \vspace{0cm}
\caption{Time evolution of $\left| \protect\rho _{eg}\right| ^{2}$ for
initial state (\ref{rho(0)term}) and $c_{e}=c_{g}=1/\protect\sqrt{2}$.
Following Ref. \protect\cite{Gleyzes}, $\protect\omega =51.099$ GHz and $g=2%
\protect\pi \times 50$ kHz. }
\label{figT}
\end{figure}

In this section, we assume that the field mode is previously thermalized:
\begin{equation}
\rho \left( 0\right) =\left( c_{e}\left| e\right\rangle +c_{g}\left|
g\right\rangle \right) \left( c_{e}^{\ast }\left\langle e\right|
+c_{g}^{\ast }\left\langle g\right| \right) \otimes \left( 1-e^{-\frac{\hbar
\omega }{k_{B}T}}\right) \underset{n=0}{\overset{\infty }{\sum }}e^{-\frac{%
\hbar \omega n}{k_{B}T}}\left| n\right\rangle \left\langle n\right| ,
\label{rho(0)term}
\end{equation}
where $k_{B}$ is Boltzmann's constant and $T$ is the absolute temperature 
\cite{Kubo}.
In Fig. \ref{figT}, we show the evolution of $\left| \rho _{eg}\right| ^{2}$ 
taking into account the same parameters
as in the experiment described in Ref. \cite
{Gleyzes}. We see that the higher the temperature of the initial state, the
more the bosonic system is effective to destroy $\left| \rho _{eg}\right| $ in the long time regime. However,
higher temperatures also retard this regime. These results are analogous to those found in 
Sections \ref{AR} and \ref{SI}.
In the actual experimental setup, the field decays according to 
a cavity damping time 
$T_{c}=130$ ms (what corresponds to $gt=4.08\times
10^{4}$). Since $T_{c}$ is much longer than the time spent to achieve
the final regime observed, the imperfections of the cavity were disregarded. 
For the actual temperature of the experimental setup $0.8$ K,
the field is near the vacuum and it occurs almost complete cyclical
recurrence of $\left| \rho _{eg}\right|$. Nevertheless, for the other temperatures in Fig. \ref
{figT}, interference fringes will be hardly observed if atom and field had interacted for a long time.

\section{Conclusions}

\label{C}

Short-time power series expansions are usually employed to define characteristic time scales for different physical quantities. We show that such a time scale is not the proper tool to analyse the fringe contrast decay due to long atom field interaction times in a Ramsey interferometer as the one described in Ref. \cite{Bertet}. This can be understood by observing that, for this model, higher order terms become relevant very early in the scale defined by the second order term. For the present case, the suitable time scale is defined through the analysis of dephasing of complex phases. Of course, time scales defined by means of short time power series expansions are applicable in several situations, as, for example, in the case described in Ref. \cite{Oliveira1}, where analytical and numerical investigations show the relevance of the second order time scale for decoherence process.

According to our analytical and numerical investigations, one oscillator 
performing a dynamics with no chaotic (or chaotic like) behavior 
is capable of destroying interference fringes 
in an effectively irreversible fashion analogously to a reservoir.
However, important differences between the case studied here and 
other situations involving actual reservoirs remain.
The single oscillator can not produce the total dissipation of the
atomic energy, as some infinite degrees of freedom reservoirs do. Also, the
bosonic system will not behave as a reservoir in the sense that it relaxes
to a unique thermal equilibrium state. Another relevant difference:
as it is shown in Ref. \cite{Morigi}, the dynamics of the
spin-boson system can be reversed by applying a transformation on the atomic
part. It is also important to
stress that complete disappearing of interference fringes is not expected, but
only attenuation. As in the system studied in Ref. \cite{Oliveira2}, quantum
phenomena disappear only when we consider the finite experimental
resolution.

In the Ramsey interferometer of Ref. \cite{Bertet}, the coherent field acts
as a beam splitter. Since it is a part of a measuring device, it must act
classically \cite{Bohr}. This classicality is achieved when the energy of
the field is increased, what leads to slower atom-field entanglement. The
quantum system (atom) and the classical system (field) must interact for the
shortest time that produces the atom's state splitting required for the
interferometry. If this interaction time is long, quantum and classical
systems get entangled. 
Then, another kind of classical limit may be envisaged: the field acts as the environment,
destroying quantum coherences of the atom. The crucial factor seems not to be the energy 
of the field, but the photon number variance.

\acknowledgements{We are grateful to M. C. Nemes, J. G. Peixoto de Faria, and R. Rossi Jr. for fruitful discussions. 
A. R. B. M. acknowledge FAPEMIG, CNPq and CEFET-MG for partial financial support. A. C. O. acknowledge FAPESB 
for partial financial support.}

\newpage

\newpage

\newpage

\newpage

\newpage

\newpage

\end{document}